\documentclass{article}

\usepackage{microtype}
\usepackage{graphicx}
\usepackage{subfigure}
\usepackage{booktabs} % for professional tables
\usepackage{color}
\usepackage{times}
\usepackage{soul}
\usepackage{makecell}
\usepackage{hyperref}

\usepackage[accepted]{icml2024}

\usepackage{amsmath}
\usepackage{amssymb}
\usepackage{mathtools}
\usepackage{amsthm}

\usepackage[capitalize,noabbrev]{cleveref}

\theoremstyle{plain}

\theoremstyle{definition}

\theoremstyle{remark}

\usepackage[textsize=tiny]{todonotes}

\icmltitlerunning{Submission and Formatting Instructions for ICML 2024}

\begin{document}

\twocolumn[
\icmltitle{Single-Cell RNA-seq Synthesis with Latent Diffusion Model}



\icmlsetsymbol{equal}{*}

\begin{icmlauthorlist}
\icmlauthor{Yixuan Wang}{equal,sch}
\icmlauthor{Shuangyin Li}{equal,sch}
\icmlauthor{Shimin DI}{yyy}
\icmlauthor{Lei Chen}{comp}

\end{icmlauthorlist}

\icmlaffiliation{yyy}{Department of CSE, HKUST, Hongkong, China}
\icmlaffiliation{comp}{Department of CSE, HKUST DSA Thrust. HKUST(GZ), Guangzhou, China}
\icmlaffiliation{sch}{School of Computer Science, South China Normal University, Guangzhou, China}

\icmlcorrespondingauthor{Yixuan Wang}{yixuanwang@m.scnu.edu.cn}
\icmlcorrespondingauthor{Shuangyin Li}{shuangyinli@scnu.edu.cn}

\icmlkeywords{Machine Learning, ICML}

\vskip 0.3in
]



\printAffiliationsAndNotice{\icmlEqualContribution} % otherwise use the standard text.

\begin{abstract}
The single-cell RNA sequencing (scRNA-seq) technology enables researchers to study complex biological systems and diseases with high resolution. The central challenge is synthesizing enough scRNA-seq samples; insufficient samples can impede downstream analysis and reproducibility. While various methods have been attempted in past research, the resulting scRNA-seq samples were often of poor quality or limited in terms of useful specific cell subpopulations. To address these issues, we propose a novel method called Single-Cell Latent Diffusion (SCLD) based on the Diffusion Model. This method is capable of synthesizing large-scale, high-quality scRNA-seq samples, including both 'holistic' or targeted specific cellular subpopulations within a unified framework. A pre-guidance mechanism is designed for synthesizing specific cellular subpopulations, while a post-guidance mechanism aims to enhance the quality of scRNA-seq samples. The SCLD can synthesize large-scale and high-quality scRNA-seq samples for various downstream tasks. Our experimental results demonstrate state-of-the-art performance in cell classification and data distribution distances when evaluated on two scRNA-seq benchmarks. Additionally, visualization experiments show the SCLD's capability in synthesizing specific cellular subpopulations.
\end{abstract}

\section{Introduction}
% Biological systems are usually highly complex, as the communication is coordinated through nonlinear interactions of tens to hundreds of thousands of different molecules between intracellular and intercellular \cite{par1}. In order to study and analyze the underlying information in these systems, researchers often rely on cell sequencing techniques. However, traditional sequencing methods only measure the average signal in a group of cells, which lead to masking cell heterogeneity and rare populations. The scRNA-seq technology is a powerful tool to amplify and extract small amounts of RNA from individual cells. In 2013, this technology was highlighted as the "Method of the Year" \cite{method}. Its ability to deepen our understanding of complex biological systems has created numerous applications in the development of human health technologies. For example, scRNA-seq in the immune system has been used to discover new immune cell populations \cite{single}, targets, and relationships, lead further bringing forward the attempt at new treatments.
Biological systems are typically characterized by their high complexity, as they involve the coordinated communication between tens to hundreds of thousands of different molecules through nonlinear interactions within cells (intracellular) and between cells (intercellular) \cite{par1}.  This intricate network of interactions plays a crucial role in various physiological and pathological processes, making it an essential area of study for understanding and developing novel therapeutic approaches. To study and analyze the underlying information in these complex systems, researchers often rely on cell sequencing techniques that offer insights into the molecular profiles of cells.  However, traditional sequencing methods, such as bulk RNA sequencing, measure the average signal in a group of cells.  This approach can lead to masking of cell heterogeneity and obscure the presence of rare cell populations, which may have significant functional implications in biological processes. The advent of single-cell RNA sequencing (scRNA-seq) technology has revolutionized the field of cellular research by enabling the amplification and extraction of small amounts of RNA from individual cells.  In 2013, this groundbreaking technology was recognized as the "Method of the Year" due to its potential to significantly enhance our understanding of complex biological systems \cite{method}.  scRNA-seq has since proven to be a powerful tool that provides unprecedented resolution at the single-cell level, revealing cell-to-cell variability and identifying rare cell types that may have been otherwise overlooked.

\begin{figure}[t]
\centering
\includegraphics[width=0.5\textwidth]{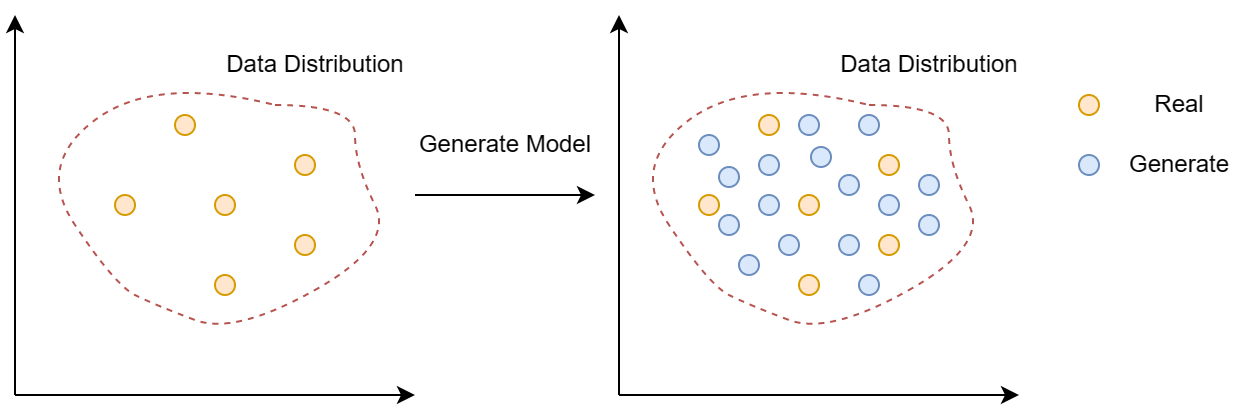}
\caption{Example of data distribution with a small quantity of data (left), and the distribution example generated after generating related data through a generative model (right).}
\label{fig:twopics}
\end{figure}

Even though the scRNA-seq technology provides excellent convenience for researchers, unfortunately, during the application of it, a limiting factor still remains: only a small number of patient scRNA-seq samples are available, which may be related to financial, ethical, or patient availability. The small scRNA-seq samples size may not accurately reflect the overall situation and may prevent other researchers from obtaining the same experimental results due to potential deviation \cite{par2}.  Even though large and well-funded projects have generated the Human Cell Atlas \cite{human} and the Mouse Cell Atlas \cite{mouse}, they are limited to broad overviews of cell populations in these tissues and organs. While the number of biological scRNA-seq samples is too limited, the current realistic in silico generation of observations can minimize this unfavorable situation \cite{standard}. The generation of a sufficient number of scRNA-seq samples in computational genomics is an urgent task and still remains formidable challenges.

Totally, synthesizing a large scale of scRNA-seq data still faces two main challenges: (1) synthesizing more realistic scRNA-seq samples, as marked dissimilarity from real scRNA-seq samples, can diminish the trustworthiness of experimental conclusions drawn from these data, and (2) the synthesis of specific cell subpopulations remains limited in quantity. Previous approaches, which typically use a single framework to synthetize all cells and then use a separate module to extract specific cell subpopulations, or train a separate model specifically, are unsatisfactory. Methods based on Generative Adversarial Networks (GAN) \cite{GAN} tend to suffer from mode collapse, where the generator produces only a limited range of outputs, and the methods based on Variational Autoencoder (VAE) \cite{GAN} are difficult to synthesize high-quality scRNA-seq samples, particularly in complex data distributions. They both fail to address the aforementioned issues. Therefore, it is an urgent that the issues of model-generated scRNA-seq samples quality and synthesizing specific cell subpopulations should be urgently addressed. 

Thus, in this paper, we propose a novel Single-Cell Latent Diffusion (SCLD), a diffusion model-based approach, to address these issues \cite{beat} (Figure \ref{fig:twopics}). Based on the capacity of the diffusion model, the SCLD can be proven effective in addressing challenges such as synthesized scRNA-seq samples quality, and the large-scale specific cell subpopulations can be synthesized through training guided by category information. For challenge (1), the SCLD uses a well-designed post-guidance mechanism to assign scores to scRNA-seq samples and outputs the high-scoring scRNA-seq samples. With the strong synthesis ability of the diffusion model and the post-guidance mechanism, the SCLD can make a significant improvement in the quality of the synthesized scRNA-seq samples. For challenge (2), the SCLD uses a tailored pre-guidance mechanism, and the average expression of different category cells and all cells is used as guiding information for the diffusion model to train the model jointly. Ultimately, the SCLD learns the features of different cells and all cells, synthesizing either “holistic” or specific cellular subpopulations in a single framework. In summary, our work makes the following \textbf{contributions}:
\begin{enumerate}
	\item To the best of our knowledge, the proposed SCLD is the first work to successfully explore the diffusion model to the single-cell sequencing task in computational genomics, by which, more realistic and flexible scRNA-seq samples can be obtained.
	\item A pre-guidance mechanism and a post-guidance mechanism are designed for genomics data. The pre-guidance mechanism uses clustering information of cells to guide model training and enables condition-based synthesis. The post-guidance mechanism operates after synthesis to produce higher-quality cells, improving model performance.
	\item The SCLD can synthetize either “holistic” or specific cellular subpopulations in a single framework, eliminating the need for separate training of two models or the additional training of a cell classifier.
\end{enumerate}

% We demonstrate superiority in the SCLD by achieving state-of-the-art results on the public datasets (PBMC \cite{pbmc} and Brain Small) in comparison to prior work, and the ability to synthetize specific cellular subpopulations is demonstrated through visualization experiments.

In this paper, we present the Single-Cell Latent Diffusion (SCLD) model, a novel approach for generating high-quality single-cell RNA sequencing (scRNA-seq) data.  We demonstrate the superiority of the SCLD model by achieving state-of-the-art results on two public datasets, PBMC \cite{pbmc} and Brain Small, in comparison to prior work.  
% Our approach is tailored for the data generation task, where the goal is to generate realistic scRNA-seq samples that resemble true biological samples.  We evaluate the performance of our model using Area Under the Curve (AUC) and Maximum Mean Discrepancy (MMD) metrics, showcasing the model's ability to generate high-quality samples that closely resemble the original data.
% Furthermore, we demonstrate the ability of our model to synthesize specific cellular subpopulations through visualization experiments.  
% In the interest of transparency and reproducibility, we have made our code and data publicly available.  This will enable researchers to build upon our work and further advance the field of single-cell data generation and analysis.

\section{Related Works}
\subsection{ScRNA-seq in Biology}
In the field of medicine, researchers use single-cell gene sequencing data to study and understand various aspects of human health and disease. Single-cell gene sequencing allows researchers to amplify and extract small amounts of RNA from individual cells, enabling the investigation of cell-to-cell heterogeneity and rare cell populations. This can provide insights into the underlying mechanisms of diseases and can inform the development of new diagnostic and therapeutic approaches.
For example, single-cell gene sequencing has been used to study the immune cell populations \cite{single} and the function in the context of cancer, autoimmune diseases, and infections \cite{bin1, bin2, bin3}. It has also been used to identify new cell types and subtypes and to understand the roles of different cell types in the development and progression of various diseases. Single-cell gene sequencing has also been used to study the development of organs and tissues, including the brain, heart, and liver, and to identify the specific cell subpopulations types and signaling pathways involved in normal and abnormal development.

\subsection{ScRNA-seq in Silico}
In the field of medicine, lack of scRNA-seq samples is a common challenge, which may be related to financial, ethical, or patient availability, as well as the inherent variability and rarity of specific cell subpopulations populations. To address this challenge, researchers have developed various techniques for augmenting and synthesizing single-cell gene data.
\subsubsection{Statistical Methods}
These methods involve analyzing the data using statistical techniques, such as clustering, dimensionality reduction, and regression. Statistical methods can help to identify patterns and relationships in the data and can synthesize similar cellular data, such as Splatter \cite{splatter} and SUGAR \cite{SUGAR}, which are able to capture the marginal distribution of the read counts to some extent, but they have difficulty in learning the joint distribution of these counts.
% Various methods have been developed for the analysis of single-cell RNA sequencing (scRNA-seq) data, which involve the application of statistical techniques, such as clustering, dimensionality reduction, and regression.  These statistical approaches can help identify patterns and relationships within the data, as well as synthesize similar cellular information.
% Examples of such methods include Splatter \cite{splatter} and SUGAR \cite{SUGAR}, which are capable of capturing the marginal distribution of read counts to a certain extent.  However, these methods face challenges in learning the joint distribution of these counts, which limits their ability to accurately model complex biological systems.
% By addressing these limitations in existing statistical techniques, research can develop more effective methods for the analysis and synthesis of scRNA-seq data.
\subsubsection{Generative Model-Based Methods}
% These methods involve using generative models, such as VAEs or GANs, to learn the underlying structure of the data and generate synthetic data. Generative models can be trained by using the available single-cell gene sequencing data and can then be used to synthesize new data that reflects the patterns and structures learned from the existing data. They can be particularly useful when working with limited data, as they can help to increase the size of the dataset and improve the performance of machine learning models.  There are several methods proposed, with both VAE and GAN having been proven to perform well in the synthesis of scRNA-seq samples, such as SC-GAN \cite{SC-GAN}, scIGANs \cite{sclGANs} and ACTIVA \cite{ACTIVA}. SC-GAN successfully applied GANs to computational biology, but the quality of the synthesized scRNA-seq samples is relatively low. A separate CSC-GAN \cite{SC-GAN} is trained in order to synthetize specific cell subpopulations. Another method, ACTIVA, uses VAE as a basis and employs a classifier as a guide to train the model, improving the quality of the scRNA-seq samples. However, generating specific cell subpopulations with ACTIVA requires filtering with a classifier, leading to the wastage of many useful scRNA-seq samples. Unfortunately, the above models do not well address the challenges presented in this field.

Generative models like VAEs and GANs are used to synthesize new single-cell gene sequencing data by learning from existing data. These models are especially beneficial for augmenting limited datasets and enhancing machine learning model performance. Notable examples include SC-GAN \cite{SC-GAN} and scIGANs \cite{sclGANs}, which leverage GANs, and ACTIVA \cite{ACTIVA}, which uses a VAE with a classifier for guidance. While SC-GAN pioneered the use of GANs in computational biology, its synthesized data quality is modest. CSC-GAN \cite{SC-GAN}, a derivative of SC-GAN, specifically focuses on generating certain cell subpopulations. ACTIVA, on the other hand, creates high-quality samples but its process of filtering with a classifier results in the loss of potentially valuable data. However, these models still fall short in fully addressing the complexities of scRNA-seq data synthesis.

\subsection{Diffusion Model}
The diffusion model \cite{DDPM, DDIM} is a generative model that can address the limitations of VAE and GAN. The diffusion model is based on the idea of gradually adding noise to an input data point over a series of time steps, allowing the model to smoothly interpolate between data points and generate diverse outputs. In contrast to VAE and GAN, the diffusion model can be trained by using maximum likelihood estimation, so as to make it more stable and easier to optimize. Compared to VAEs and GANs, the diffusion model has better sampling efficiency and is less prone to mode collapse \cite{beng}, so higher-quality scRNA-seq samples can be generated. Furthermore, using categorical information as guidance allows the diffusion model to have the ability to generate a large number of specific cell subpopulations. Thus, diffusion models can address issues as scRNA-seq samples quality and generate specific cell subpopulations, which cannot be done based on GAN and VAE models.
% The diffusion model \cite{DDPM, DDIM} is a generative approach that can overcome the limitations of Variational Autoencoders (VAEs) and Generative Adversarial Networks (GANs) in synthesizing scRNA-seq data. Based on the concept of incrementally adding noise to an input data point over a series of time steps, the diffusion model enables smooth interpolation between data points and generates diverse outputs.
% In contrast to VAEs and GANs, diffusion models employ maximum likelihood estimation for training, which renders them more stable and easier to optimize. As a result, they exhibit superior sampling efficiency and are less susceptible to mode collapse \cite{beng}. Consequently, diffusion models can generate higher-quality scRNA-seq samples compared to their VAE and GAN counterparts.
% Moreover, incorporating categorical information as guidance enables diffusion models to generate a vast array of specific cell subpopulations. This capability addresses the challenges of scRNA-seq sample quality and the generation of specific cell subpopulations, which are not adequately addressed by GAN and VAE models. Overall, diffusion models present a promising alternative for the synthesis of high-quality scRNA-seq data, offering unique advantages over traditional generative approaches.

\begin{table}
	\centering
 	\caption{Notations for the SCLD.}

	\resizebox{.99\columnwidth}{!}{
		\begin{tabular}{lrr}
			\toprule
			Notation  & \makecell[l]{Definition} \\
			\midrule
			\midrule
			$\mathcal{N}$     & \makecell[l]{Set of cells}       \\
			$\mathcal{G}$     & \makecell[l]{Set of genes}\\
			$\mathcal{T}$  &     \makecell[l]{Set of model iteration steps}   \\
			$\mathcal{K}$ & \makecell[l]{Set of cell clusters}\\
			\midrule
			$x_i \in \mathcal{N}$ & \makecell[l]{Single cell information}\\
			$x_i^j \in \mathcal{G}$ & \makecell[l]{Single gene information from single cell}\\
            $x_{(t)}, z_{(t)} \in \mathcal{T}$ & \makecell[l]{The sample is at the step of the model}\\
			$c_k \in \mathcal{K}$ & \makecell[l]{Specific cell subpopulations cluster}\\
			
			\midrule
			$W$  & \makecell[l]{Learnable parameterized matrices}\\
			$b$  & \makecell[l]{Learnable bias}\\
			\midrule
			$z$ & \makecell[l]{Representation of scRNA-seq samples in latent space}\\ 
			$louvain(x)$ & \makecell[l]{A clustering function that assigns categories to cells}\\
			$Round(x)$ & \makecell[l]{A function to compress scRNA-seq samples}\\
			$Map(z)$ &  \makecell[l]{A function to expansion latent space data}\\
			\midrule
			
                $\mathcal{L}_{DM}$ & \makecell[l]{A Loss function of Diffusion Model}\\
                $\mathcal{L}_{AE}$ & \makecell[l]{A Loss function of Autoencoder}\\
                $\mathcal{L}_{LD}$ & \makecell[l]{A Loss function of Latent Diffusion Model}\\
                \bottomrule
		\end{tabular}
	}
	\label{tab:notation}
\end{table}
\section{Single-Cell Latent Diffusion}
In this section, we first define the notation used in the SCLD, and then provide a brief overview of its foundation, the diffusion model. Next, we describe the modifications to adapt the diffusion model for single-cell RNA sequencing tasks. There are the pre-guidance mechanism and the post-guidance mechanism, which are vital components of the SCLD. Finally, we present the training algorithm and sampling algorithm of the SCLD.

\begin{figure*}
\begin{center}
\includegraphics[width=\linewidth]{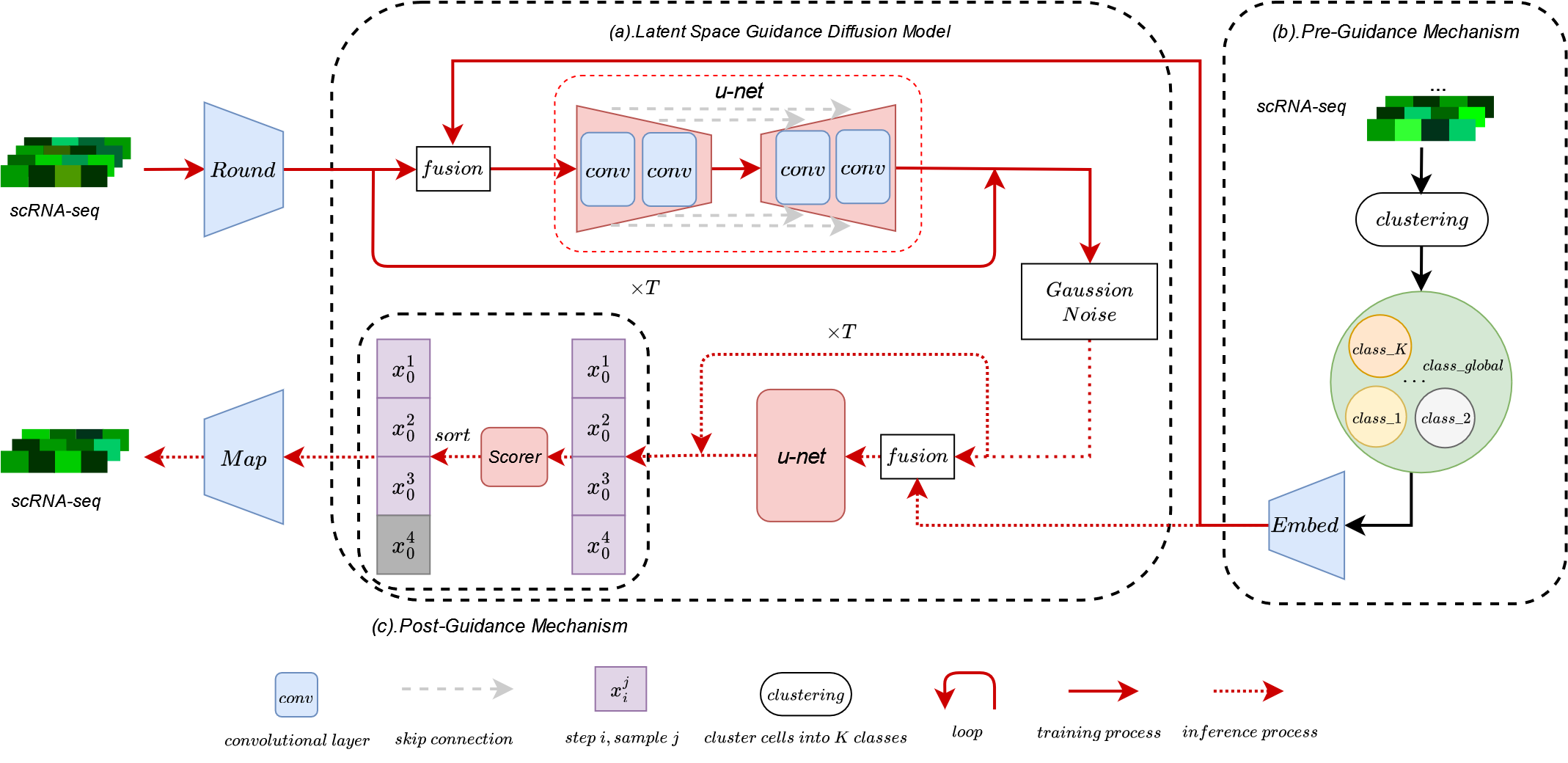}
\caption{Overview of the proposed SCLD for scRNA-seq samples synthesis, including the process of training and inference. Part (a) represents the latent diffusion model, which primarily employs a U-Net for model training. Part (b) illustrates the pre-guidance mechanism, which introduces different category information into the diffusion model. Part (c) depicts the post-guidance mechanism, which uses a scoring system to rank the generated samples and discard those of lower quality.} \label{fig:framework}
\end{center}
\end{figure*}

\subsection{Notations}
% Consider a cell $x_i \in \mathcal{N}$ has a sequence of genes ${x_{i}^{(1)},...,x_i^{(g)} }, g\in \mathcal{G}$, where each $x_i^{(g)}$ contains a positive integer gene expression value. In this paper, a category information $c_k \in \mathcal{K}$  is assigned to $x_i$ using the function $louvain$ \cite{louvain}, so that the category corresponding to $x_i$ is $c_k$. Then,  $x_i$ is compressed into the latent space $z_i$ using the function $Round$ (This function is thoroughly described in Section \ref{method}). $z_{(t)}$ will iterate for $t\in\mathcal{T}$ steps, and the expression $z$ of the synthesized cells in the latent space will eventually be output. After that $z$ is mapped to a high-dimensional space using the function $Map$ (See Section \ref{method}), and eventually reverted back to the high-dimensional representation $x$.
Consider a cell $x_i \in \mathcal{N}$ with a sequence of genes ${x_{i}^{(1)},...,x_i^{(g)} }, g\in \mathcal{G}$, where each $x_i^{(g)}$ contains a positive integer representing the gene expression value. In this paper, a category information $c_k \in \mathcal{K}$ is assigned to $x_i$ using the Louvain clustering function $louvain$ \cite{louvain}, so that the category corresponding to $x_i$ is $c_k$. Then, $x_i$ is compressed into the latent space $z_i$ using the function $Round$ (This function is thoroughly described in Section \ref{method}). $z_{(t)}$ will iterate for $t\in\mathcal{T}$ steps, and the expression $z$ of the synthesized cells in the latent space will eventually be output. After that, $z$ is mapped back to a high-dimensional space using the function $Map$ (See Section \ref{method}), and finally transformed back to the high-dimensional representation $x$.

% In our proposed model, we utilize several learnable parameters, including parameterized matrices $W$ and bias terms $b$. The loss functions $\mathcal{L}{DM}$, $\mathcal{L}{AE}$, and $\mathcal{L}_{LD}$ are employed for optimizing the Diffusion Model, Autoencoder, and Latent Diffusion Model, respectively. These loss functions are minimized during the training process to ensure the generated scRNA-seq samples are of high quality and representative of the original data.

The notations presented in Table \ref{tab:notation} provide a comprehensive summary of the key symbols used in our proposed SCLD model. By using these notations, we can effectively describe the model architecture, input and output representations, and the optimization process.

\subsection{The Proposed Method}\label{method}
The traditional diffusion models are probabilistic models that learn a data distribution $p(x)$ by iteratively denoising a normally distributed variable. This process corresponds to learning the reverse process of a fixed Markov Chain of length T. For generating fields, the most successful model uses a variant of the variational lower bound on $p(x)$ that is based on denoising score-matching. These models can be regarded as a sequence of equally weighted denoising autoencoder $t \sim \mathcal{T}$, which are trained to predict a denoised version of their input $x_{(t)}$, where $x_{(t)}$ is a noisy version of the input $x$. The loss function for these models(represented $\mathcal{L}_{DM}$) can be simplified to
\begin{align}
	\mathcal{L}_{DM} = \mathbb{E}_{t,x} || \mu - \mu_\theta(x_{(t)}, t)||^2,
\end{align}%
with t uniformly sampled from $\mathcal{T}$.

Based on the process of diffusion model, the proposed SCLD is composed of the pre-guidance mechanism and the post-guidance mechanism. During the process of inference, the input to the model is the desired specific cell subpopulations category to be generated, and the output is a specific cell subpopulations synthesized by the SCLD. The Figure \ref{fig:test} shows the process of the SCLD.

\subsubsection{Latent Space Guided Diffusion for Single-Cell}
Single-cell sequencing data is typically bag-of-words \cite{bow} data because there is no upper limit to the value of gene expression per cell and the order between genes does not affect the cell. Bag-of-words are discrete data, which can be expressed as shown in Eq. (\ref{for:number}). The diffusion model, which specifically deals with continuous data, is not ideal for processing discrete data. A Generative Model of Latent Representations with an autoencoder can map scRNA-seq samples into a continuous latent space, allowing the diffusion model to be trained. In addition, it can access an efficient, low-dimensional latent space in which high-frequency, imperceptible details are abstracted away. In contrast to the original data space, the latent space is more suitable for the SCLD in this task, as it can now (i) train with continuous data, (ii) focus on the important gene information of the data, and (iii) train in a low-dimensional space with higher computing efficiency.
\begin{align}
\label{for:number}
  x_{i}^j = N, \; N\in \mathbf{R}.
\end{align}%

This mapping process is called $Round$. Then the diffusion can complete the process in this latent space. In the reverse process, scRNA-seq samples are mapped to the original space using an autoencoder. This process is called $Map$.

\begin{align}
  Round(x) = W_xx + b_x,\\
  Map(z) = W_zz+b_z.
\end{align}%

In detail, there is a pair of symmetrics, a two-layer linear encoder and a decoder and an activation function is used behind each layer, which map high-dimensional data to low dimensions. Its loss function(represented $\mathcal{L}_{AE}$) is the Mean Squared Error (MSE) by measuring the gap between $x$ and $x^{'}$ in $\mathcal{L}_{AE}$.
 
\begin{align}
  z_i = Round(x_i), 
  x^{'}_i = Map(z_i)\; x_i \in \mathcal{N},
\end{align}%
\begin{align}
  \mathcal{L}_{AE}(x)=\sum_{i=1}^N||x_i - x^{'}_i||^2.
\end{align}%

Therefore, the loss function of the above diffusion model(represented $\mathcal{L}_{LD}$) needs to be rewritten into the loss of the potential diffusion model:
\begin{align}
 \mathcal{L}_{LD}(x_0) = \mathbb{E}_{t,x} || \mu- \mu_\theta(z_{(t)}, t)||^2.
\end{align}%

\begin{figure*}[h]
\begin{center}
\includegraphics[width=\linewidth]{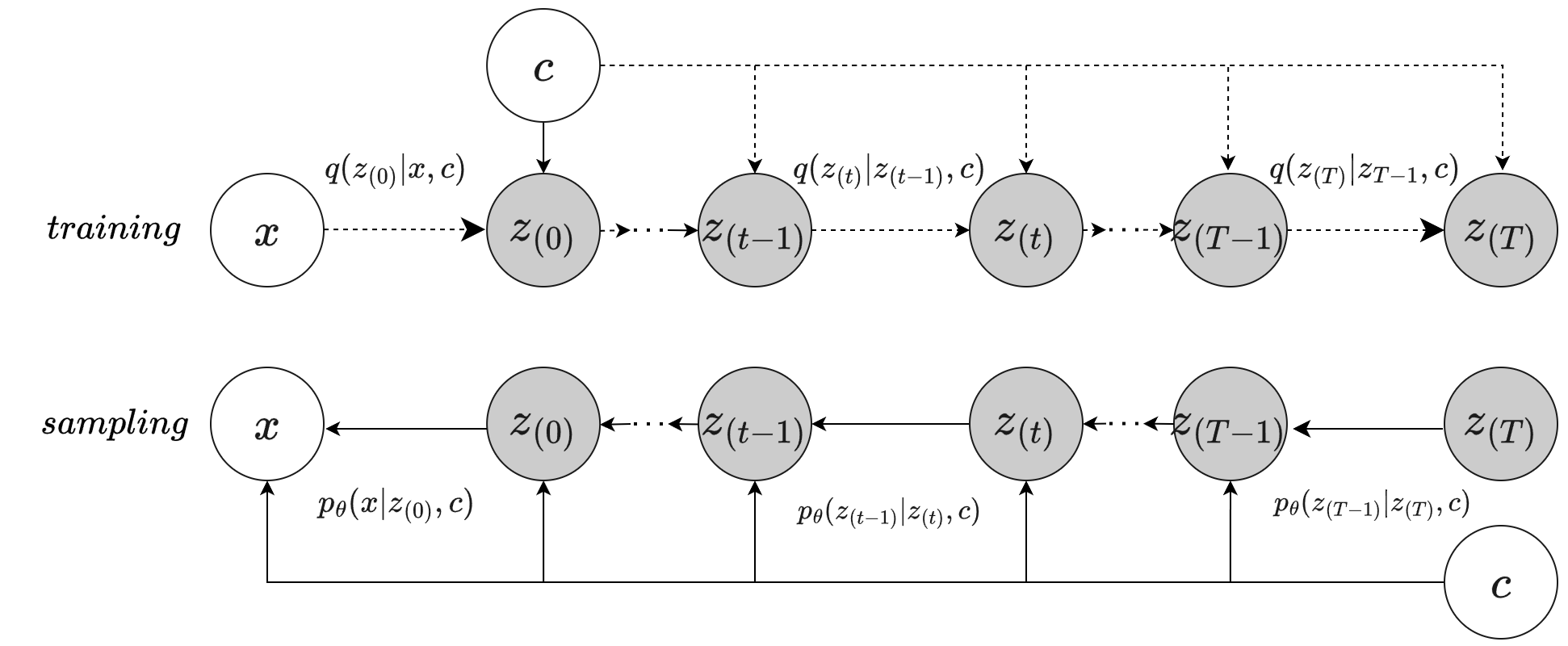}
\caption{The directed graphical model considered in the SCLD. $q(z_{(0)}|x, c)$, $p_\theta(z_{(T-1)}|z_{(T)}, c)$ represent noising adding forward process and denoising backward process, respectively.} \label{fig:graph}
\end{center}
\end{figure*}

\subsubsection{Pre-Guidance Mechanism}
As mentioned above, scRNA-seq samples are unimodal \cite{dan}, and lack the important factor in scRNA-seq samples generation, the category information. To address this issue, we propose a pre-guided mechanism to extract category information from the data and inject it into the diffusion model, which also helps to generate higher-quality data.

In detail, first, the graph-based clustering algorithm louvain is used to cluster all cells into $K$ categories. Then an encoder maps all cells in each cluster to their average expression as $c_k$, and maps the genes of all cells to their average expression, using it as global information expressed by $c_{K+1}$, and can be expressed by the following Eq.:
% add some words
\begin{align}
   c_k =\left\{\begin{aligned}
 &louvain(x_i), \;x_i\in \mathcal{N}\\
&\sum_{i=0}^{N}x_i,\;x_i\in\mathcal{N}
\end{aligned}
, c_k \in \mathcal{K}
\right..
\end{align}%

Then the SCLD obtains embeddings from all cluster information through a Multilayer Perceptron \cite{MLP} encoder named $Embed$, and trains with concatenated original data and embeddings. It is worth mentioning that, in the fusion operation, there are other fusion methods (e.g. adding and cross attention), but concatenation still performes the best.

To train this diffusion model, we define a forward process in which each transition from $x_{(t-1)}$ to $x_{(t)}$ can be parametrized by 
\begin{align}
q(z_{(t)} | z_{(t-1)},c) = \mathcal{N} (z_{(t)}; \sqrt{1 - \beta_t }(z_{(t-1)}+c), \sqrt{\beta_t}I), 
\end{align}
where $\beta_t$ is a hyperparameter representing the amount of noise added to the scRNA-seq samples at step $t$. The training objective of the forward process involves generating noisy data according to a pre-defined forward process $q$ and the data predicted by the model.

Originally, the goal is to maximize the marginal likelihood of the data $ \mathbb{E}_{x_{(0)} \sim p_{data}} [log p_\theta (x_{(0)},c)]$, and thus, the loss function of the diffusion model(represented $\mathcal{L}$) can be derived \cite{15},
\begin{align}
	\mathcal{L}(z_{(0)},c) = & \mathbb{E}_{q_{(z_{(1:T)}|z_{(0)},c)}}[log\frac{(q(z_{(T)}|z_{(0)},c)}{p_\theta(z_{(T)},c)} 
    \\ 
    + \sum_{t=2}^T &log\frac{q(z_{(t-1)}|z_{(0)}, z_{(t)},c)}{p\theta(z_{(t-1)}|z_{(t)},c)} \nonumber
	-   logp_{\theta}(z_{(0)}|z_{(1)},c)].
\end{align}%

Following the original DDPM parameterization, and we make corresponding changes to $\mathcal{L}$ based on the definition:
\begin{align}
    \mathcal{L}(z_{(0)},c) = \mathbb{E}_{t,z,c} || \mu - \mu_\theta(z_{(t)}, c, t)||^2,
\end{align}
where $\mu$ is the mean of the posterior $q(x_{(t-1)}|z_{(0)}, z_{(t)},c)$, which is a closed-form Gaussian, and $\mu_\theta(x_t, t)$ is the predicted mean of $p_\theta(z_{(t-1)} | z_{(t)},c)$ computed by a U-Net \cite{unet} (similar to prior work). While $\mathcal{L}$ is no longer a valid lower bound, previous work has proved that training with this goal can produce higher quality scRNA-seq samples and the process is stable \cite{stable}.

\subsubsection{Post-Guidance Mechanism}
In computational genomics, high-quality scRNA-seq samples are significant to ensure researchers can conduct downstream analysis effectively. Otherwise, researchers may obtain incorrect patterns and conclusions with them, which is entirely unacceptable.

To address this issue, we propose a post-guidance mechanism to improve scRNA-seq samples quality by assigning and ranking. There is a $Scorer$, which operates after that the diffusion model completes sampling. The $Scorer$ assigns points to generated scRNA-seq samples, and sets a threshold value, keeping scRNA-seq samples that reach the threshold while discarding others. This greatly improves the confidence of generated cells.

In detail, scRNA-seq samples produced by the diffusion model are transmitted to the $Scorer$ for scoring. The $Scorer$ is composed of three linear layers and produces a two-dimensional output mapping to the scores of real and fake cells, which are used to train with the loss function with binary cross entropy.

Each scRNA-seq sample will receive a corresponding score, and then all scRNA-seq samples are sorted by the scores. The top 80$\%$ of them are used as the final outputs of the model (If the threshold is set too low, it will lead to high reasoning consumption, but if it is set too high, the effect of the post-guidance mechanism will not be obvious. After weighing efficiency and effectiveness, we set it to 80$\%$.). This mechanism has been shown to significantly improve the quality of scRNA-seq samples generation in the experiments.

After the training process, the process from $x_{(t)}$ to $x_{(0)}$, each denoising step is parametrized by 
\begin{align}
    p_\theta(z_{(t-1)}|z_{(t)}, k) = 
    \mathcal{N} (z_{(t-1)}; \mu_\theta(z_{(t)}, t), \Sigma_\theta(z_{(t)}, t)),
\end{align}
where $\mu_\theta$ and $\Sigma_\theta$ are fitted by a U-Net. In the sampling process, classifier-free guidance \cite{classifier} is used because different clusters of cells and all cells of specific cell subpopulations are used in the pre-guidance mechanism for joint training, so the Eq. (\ref{e}) should be used to carry out, which has been proven effective,

\begin{align}
\label{e}
 \Tilde{\epsilon}_\theta(z, c_i)=(1+\omega)\epsilon_\theta(z,c_i)-\omega\epsilon_\theta(z, c_{K+1}).
\end{align}%

Note that $\omega$ is the guidance weight, setting $\omega$ = -1 disables classifier-free guidance, while increasing $\omega \geq$ -1 strengthens the effect of guidance \cite{imagen}.  Figure \ref{fig:graph} presents the entire process of SCLD in more detail, in the form of a graph structure.

\begin{table}
    \centering
        \caption{Statistics of the datasets.}

    \begin{tabular}{lllll}
        \hline
        Dataset  & Tissue & Cells & Genes & Clusters \\
        \hline
        PBMC & Blood & 68,579 & 17,789 & \multicolumn{1}{@{}c@{}}{10}\\
        Brain Small & Brain & 20,000 & 17,970 & \multicolumn{1}{@{}c@{}}{8}\\
        \hline
    \end{tabular}
    \label{tab:datasets}
\end{table}

\section{Experiments}

% \subsection{Experiments Setting}
\subsection{Datasets}
% The SCLD is verified by the benchmark datasets, and compared with prior work. All experiments are implemented on 68k PBMC and Brain Small. \\
The Single-Cell Latent Diffusion (SCLD) model was rigorously tested using two benchmark datasets, 68k PBMC and Brain Small (Table~\ref{tab:datasets}), and compared with previous studies.

\begin{table}
    \centering
    \caption{The results on two scRNA-seq benchmark datasets (\dag: adding pre-guidance mechanism, and \ddag: adding post-guidance mechanism.).}
    \resizebox{.99\columnwidth}{!}{
    \begin{tabular}{lrrrr}
        \toprule
        & \multicolumn{2}{@{}c@{}}{68k PBMC} & \multicolumn{2}{@{}c@{}}{Brain Small} \\
        \cmidrule{2-3}\cmidrule{4-5}%
        Method  & \multicolumn{1}{@{}c@{}}{AUC$\downarrow$} & \multicolumn{1}{@{}c@{}}{MMD$\downarrow$}& \multicolumn{1}{@{}c@{}}{AUC$\downarrow$} & \multicolumn{1}{@{}c@{}}{MMD$\downarrow$}\\
        \midrule
        positive    &   0.51$_{\pm0.01}$    &  0.05$_{\pm0.01}$    
                &    0.50$_{\pm0.01}$   &  0.02$_{\pm0.01}$  \\
        \midrule
        Random & 1.00$_{\pm0.00}$ & \multicolumn{1}{@{}c@{}}{-}
        & 1.00$_{\pm0.00}$ &\multicolumn{1}{@{}c@{}}{-}\\
        Splatter& 1.00$_{\pm0.00}$ &129.52$_{\pm0.01}$ 
        &\multicolumn{1}{@{}c@{}}{-} &\multicolumn{1}{@{}c@{}}{-} \\
        
        SUGAR &0.98$_{\pm0.01}$  &59.45$_{\pm0.01}$  
        &\multicolumn{1}{@{}c@{}}{-} & \multicolumn{1}{@{}c@{}}{-}\\
        
        SC-GAN     & 0.73$_{\pm0.01}$          &  0.90$_{\pm0.01}$      
                  & 0.74$_{\pm0.01}$           & 0.97$_{\pm0.01}$         \\
        ACTIVA  & 0.68$_{\pm0.01}$          &   0.80$_{\pm0.01}$      
                & 0.62$_{\pm0.01}$          &   0.77$_{\pm0.01}$\\
        \midrule
        $SCLD$ &     0.65$_{\pm0.01}$      &    0.77$_{\pm0.01}$     
                  &    0.68$_{\pm0.01}$       & 0.82$_{\pm0.01}$ \\
        $SCLD^{\dag}$  & 0.64$_{\pm0.01}$  &   0.75$_{\pm0.01}$
                & 0.68$_{\pm0.01}$  &   0.81$_{\pm0.01}$    \\
        $SCLD^{\ddag}$ & \textbf{0.59}$_{\pm0.01}$ &    \textbf{0.65}$_{\pm0.01}$
                & \textbf{0.61}$_{\pm0.01}$  &   \textbf{0.72}$_{\pm0.01}$    \\
        \bottomrule
    \end{tabular}
    }
    
    \label{tab:results}

\end{table}

% \begin{figure*}[ht]
% \centering
% \subfigure[$PBMC(AUC)$]{
% \begin{minipage}[t]{0.49\linewidth}
% \centering
% \includegraphics[width=\linewidth]{pic/res1.png}
% %\caption{fig1}
% \end{minipage}%
% }%
% \subfigure[$Brain\ Small(AUC)$]{
% \begin{minipage}[t]{0.49\linewidth}
% \centering
% \includegraphics[width=\linewidth]{pic/res2.png}
% %\caption{fig1}
% \end{minipage}%
% }%

% \subfigure[$PBMC(MMD)$]{
% \begin{minipage}[t]{0.49\linewidth}
% \centering
% \includegraphics[width=\linewidth]{pic/res3.png}
% %\caption{fig1}
% \end{minipage}%
% }%
% \subfigure[$Brain\ Small(MMD)$]{
% \begin{minipage}[t]{0.49\linewidth}
% \centering
% \includegraphics[width=\linewidth]{pic/res4.png}
% %\caption{fig1}
% \end{minipage}%
% }%
% \caption{The results on two scRNA-seq benchmark datasets (SCLD+pre: adding pre-guidance mechanism, and SCLD+pre+post: adding pre-guidance and post-guidance mechanism.)}
% \label{fig:result}
% \end{figure*}

\begin{figure}[ht]
\begin{center}
\includegraphics[width=0.8\linewidth]{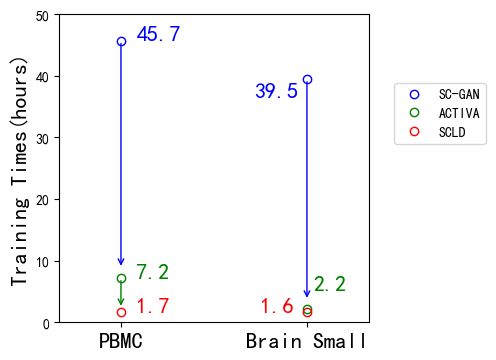}
\caption{Average training time with scRNA-seq samples standard deviation (in hours) for the SCLD, ACTIVA and SC-GAN in the same computational environment.} \label{fig:time}
\end{center}
\end{figure}

% todo words

% \begin{figure}
% \begin{center}
% \includegraphics[width=0.8\linewidth]{pic/alb.png}
% \caption{Ablation experiments on the fusion operation of category information and the SCLD.} \label{fig:alb}
% \end{center}
% \end{figure}

% \begin{table}
%     \centering
%     \caption{ Average training time with scRNA-seq samples standard deviation (in seconds) for the SCLD, ACTIVA and SC-GAN in the same computational environment.}
%     \begin{tabular}{lrrrr}
%         \toprule
%         & \multicolumn{2}{@{}c@{}}{68k PBMC} & \multicolumn{2}{@{}c@{}}{Brain Small} \\
%         \cmidrule{2-3}\cmidrule{4-5}%
%         Method  & \multicolumn{1}{@{}c@{}}{seconds} & \multicolumn{1}{@{}c@{}}{hours}& \multicolumn{1}{@{}c@{}}{seconds} & \multicolumn{1}{@{}c@{}}{hours}\\
%         \midrule
%         SC-GAN     & 164839$_{\pm1}$           &  45.7   
%                   & 142238$_{\pm1}$           &  39.5       \\
%         ACTIVA  &  26025$_{\pm1}$       &   7.2      
%                 & 8074$_{\pm1}$          &   2.2\\
%         \midrule
%         $SCLD^{\ddag}$ & 5974$_{\pm1}$    & 1.7
%         & 5833$_{\pm1}$     & 1.6\\
%         \bottomrule
%     \end{tabular}

%     \label{tab:times}
% \end{table}

\begin{table}[t]
    \centering
    \caption{Ablation experiments on the fusion operation of category information and the SCLD.}

    \begin{tabular}{lrrr}
        \toprule
        Datasets & \multicolumn{1}{@{}c@{}}{Add} & \multicolumn{1}{@{}c@{}}{Cat} & \multicolumn{1}{@{}c@{}}{Atten}\\
        % \cmidrule{2-4}%
          %& \multicolumn{3}{@{}c@{}}{AUC}\\
        \midrule
        68k PBMC     & 0.64$_{\pm0.01}$          &  0.65$_{\pm0.01}$                  & 0.68$_{\pm0.01}$               \\
        Brain Small  & 0.68$_{\pm0.01}$          &   0.68$_{\pm0.01}$                  & 0.69$_{\pm0.01}$          \\
        \bottomrule
    \end{tabular}
    \label{tab:fusion}
\end{table}

% \begin{figure*}[h]
% \centering
% \subfigure{
% \begin{minipage}[t]{\linewidth}
% \includegraphics[width=0.9\linewidth]{pic/test.png}
% %\caption{fig1}
% \caption{Pearson correlation of marker genes for the SCLD generated (bottom left) and the real (upper right) data}
% \centering
% \label{fig:test1}
% \end{minipage}%
% }%

% \subfigure{
% \begin{minipage}[t]{\linewidth}
% \includegraphics[width=0.9\linewidth]{pic/test.png}
% %\caption{fig1}
% \caption{Pearson correlation of marker genes for the SCLD generated (bottom left) and the real (upper right) data}
% \centering
% \label{fig:test2}
% \end{minipage}%
% }%
% \label{fig:testx}

% \end{figure*}

% \begin{figure}[ht]
% \centering
% \subfigure[$PBMC$]{
% \begin{minipage}[t]{0.53\linewidth}
% \centering
% \includegraphics[width=\linewidth]{pic/reb1.png}
% %\caption{fig1}
% \end{minipage}%
% }%
% \subfigure[$Brain\ Small$]{
% \begin{minipage}[t]{0.53\linewidth}
% \centering
% \includegraphics[width=\linewidth]{pic/reb2.png}
% %\caption{fig1}
% \end{minipage}%
% }%
% \caption{Pearson correlation of marker genes for the SCLD generated (bottom left) and the real (upper right) data.}
% \label{fig:reb}
% \end{figure}

\subsection{Quality of ScRNA-seq Samples}
% The assessment of generative model quality is a longstanding and ongoing research area that poses significant challenges \cite{hard1,hard2}. To evaluate the synthetic scRNA-seq generated by the SCLD, both qualitative and quantitative metrics are applied, as employed in Marouf et al. and Ali Heydari. For qualitative evaluation, the experiment compares the manifolds of generated and real cells using t-SNE \cite{tsne}, as demonstrated in Section 4.3. For quantitative evaluation, a Random Forest(RF) classifier \cite{rf} is trained to distinguish between real and synthetic cells. Specifically, a portion of the synthesized scRNA-seq samples are used for training the RF classifier and the remainder for calculating the AUC, which serves as a measure of model performance. In addition to AUC, the maximum mean discrepancy (MMD) \cite{MMD} between real and synthetic cells are utilized as an evaluation metric, as MMD is a distance measure, and a lower value indicates better proximity between the two distributions.

The assessment of generative model quality is a longstanding and ongoing research area, presenting significant challenges \cite{hard1,hard2}. To evaluate the synthetic scRNA-seq data generated by the SCLD model, we applied both qualitative and quantitative metrics, as utilized by Marouf et al. and Ali Heydari.

For qualitative assessment, the experiment compares the manifolds of generated and real cells using t-SNE \cite{tsne}, as demonstrated in Section 4.3. This visualization technique enables us to assess the similarity between the distributions of real and synthetic cells in reduced-dimensional space.
For quantitative assessment, we employed a Random Forest (RF) classifier \cite{rf} to distinguish between real and synthetic cells. The RF classifier is an ensemble learning method based on constructing multiple decision trees. The classification decision is made by aggregating the results of individual trees, which reduces the risk of overfitting and enhances overall accuracy and robustness.
To assess the performance of the SCLD model in generating realistic scRNA-seq data, we trained the RF classifier using a mix of synthesized and real samples as input features. The remaining synthesized samples were reserved for testing the classifier's performance. We then calculated the Area Under the Curve (AUC) \cite{auc} for the classifier, serving as a measure of the model's performance. A higher AUC indicates that the RF classifier has difficulty distinguishing between real and synthetic cells, suggesting that the generated samples closely resemble real data.

The AUC is derived from the Receiver Operating Characteristic (ROC) curve \cite{roc}, which plots the true positive rate (TPR) against the false positive rate (FPR) at various classification thresholds \cite{under}. The AUC metric measures the entire two-dimensional area underneath the ROC curve.

The AUC value ranges from 0 to 1, where a higher AUC indicates a better model performance. An AUC of 0.5 suggests that the classifier performs no better than random chance, while an AUC of 1 implies a perfect classification. The advantage of using the AUC metric is that it is invariant to class distribution and provides a single scalar value to represent the performance of a classifier across all possible thresholds.
In our experiments, we compute the AUC to evaluate the ability of our proposed model to discriminate between real and synthetic scRNA-seq samples. when the AUC approaches 0.5, it actually indicates that our model generates high-quality samples that are challenging to differentiate from the real data.

In addition to the AUC metric, we employ the maximum mean discrepancy (MMD) \cite{MMD} to evaluate the similarity between the real and synthetic cells. MMD is a distance measure that quantifies the difference between two probability distributions. A lower MMD value indicates a better proximity between the two distributions, implying that the generated cells closely resemble their real counterparts.
By leveraging both qualitative and quantitative evaluation methods, including the RF classifier and MMD metric, we can effectively assess the performance of the SCLD model in generating synthetic scRNA-seq data that closely resembles real biological samples.

% todo douhao, objection
Specifically, as shown by the Eq. (\ref{for:mmd}), the $\sigma_i$ in $MMD$ is chosen to be $\frac{m}{2}$, $m$, $2m$, and $m$ is the median of the average distance between a point to its nearest 25 neighbors,
\begin{align}
\label{for:mmd}
MMD(x, y) = \sum_i exp(-\frac{||x-y||^2}{\sigma_i^2}), 
\end{align}
where $x$ are the real samples and $y$ are the generated samples.

To show the SCLD’s performance in generation sample quality, the SCLD compared with many other methods, where the AUC and MMD are used as the evaluation index. Table \ref{tab:results} presents the best-case evaluation results (positive, `positive" is a subset of the real dataset that is extracted as negative samples to train RF, so its AUC will be very close to 0.5.) for statistical (Random, Splatter, and Sugar) and neural network-based (SC-GAN and ACTIVA) methods, as well as the SCLD. As it can be inferred from Table \ref{tab:results}, when the MMD value is lower and the AUC approaches 0.5, it indicate that the synthesized scRNA-seq samples are more realistic and that the SCLD method performs better. The results obtained by the SCLD are clearly superior to those of the other methods. In addition, ablation experiment is implemented on the fusion method of category information and the backbone network. Several mainstream methods have been evaluated, including concatenation, addition, and cross attention. The results indicate that the best performance can be achieved using the addition operation. Taken together, these experiments demonstrate the effectiveness of the SCLD.

\begin{table}[h]
\centering
\caption{The generation performance in specific cell types in 68k PBMC with AUC. }
\resizebox{0.48\textwidth}{!}{%
\begin{tabular}{|c|c|c|c|c|}
\hline
Method & cluster 1 & cluster 2 & cluster 3 & cluster 6 \\
\hline
CSC-GAN & 0.82 & 0.72 & 0.86 & 0.75\\
ACTIVA & 0.73 & - & 0.74 & -\\
SCLD(ours) & \textbf{0.71} $\downarrow$ & \textbf{0.68} $\downarrow$ & \textbf{0.70} $\downarrow$ & \textbf{0.69} $\downarrow$\\
\hline
\end{tabular}
}
\label{tab:example}
\end{table}

\subsection{Conditional Generation}

Additionally, it is worth noting that the proposed method is not only the most effective, but also the most efficient in terms of training time. As shown in Figure \ref{fig:time}, under the same hardware conditions (Tesla v100), the training time required for the SCLD is significantly shorter than that of the other two neural network-based models. This demonstrates the cost-effectiveness of the proposed method, as it is able to achieve superior performance while requiring less training time. Overall, the results of the experiments highlight the effectiveness and efficiency of the proposed method for synthesizing high-quality single-cell sequencing data.

As is shown in Figure \ref{fig:test}, when using Algorithm \ref{alg:algorithm2} for sampling, it can control the clusters of the scRNA-seq samples by adjusting the size of $\omega$. It is worth noting that when $\omega$ is set to -1, the generation disables classifier-free guidance. This is why the SCLD can realize the ability of a model integrating random generation and conditional generation. Moreover, in order to further explore the generation ability of the SCLD and ensure the model has learned different types of information and global information through the pre-guidance mechanism, different $\omega$ is tried to represent the strength of the willingness to generate specific types of information. For example, in Figure \ref{fig:test}, to demonstrate the SCLD generation ability, we first set $c$ to 4 in PBMC to represent the fourth cell subpopulation, and set it to 2 in Brain Small to represent the second cell subpopulation. Then we linearly change $\omega$ from -1 to 1 to observe the mastery of the SCLD over global and specific information. When $\omega$ is -1, the SCLD does not use any specific information, and $\omega \geq$ -1, it biases toward generating particular cell subpopulations. Finally, we set the different number of samples to avoid crowding when generating specific cell subpopulations, and we generated 3000 samples when $\omega=-1$, the same number as the real cells being tested, 2000 samples when $\omega=0$, and 1000 samples when $\omega=1$. 

In Figure \ref{fig:test}, (a) and (f) are real cells used for testing, (g) and (l) are the generated “holistic” cells. It can learn that the real cells and generated cells (red part) are very similar, and (b)-(f) and (h)-(l) are the case of generating specific cell subpopulations. It is shown that when $c=4$ in PBMC, real cells are concentrated in the red part. The location where the samples generated in (f) and (l) are correct, and by increasing $\omega$, the relative position of the generated samples is more accurate. The same is true for the Brain Small dataset. Therefore, it can be found that the larger $\omega$, the more concentrated scRNA-seq samples. This shows that the SCLD has learned the difference in information between different cells and, accurately generates designated cells while understanding the overall specific cell subpopulations, and can also synthesize "holistic" cells. 
In Table \ref{tab:example}, we individually synthesized clusters 1, 2, 3, and 6, and conducted comparisons with previous works.  We found that the SCLD consistently demonstrated superior performance.

% Furthermore, the SCLD is capable of modeling intergene dependencies and correlations, which are characteristic of biological gene regulatory networks. To demonstrate this, we calculated the correlation and distribution of the counts for cluster-specific marker genes (Figure \ref{fig:reb}) and 100 highly variable genes between the generated.

% \vspace{-0.5em}
\section{Conclusion and Future Work}
In this work, we propose a diffusion model for single-cell sequencing data that leverages cells data clustering and the average expression of all cells as guidance information and maps it to the latent space for training, thereby reducing the training costs. During the sampling process, there is a storage queue and a $Scorer$ to enable parallel processing among modules while retaining cells with high scores, significantly improving the quality of the generated cells. The experiments demonstrate that the SCLD is not only the most effective but also the most efficient. In the future, we plan to address the slow sampling speed of the diffusion model, which is an inherent limitation of its algorithm. 
\nocite{langley00}

\bibliography{example_paper}
\bibliographystyle{icml2024}

%%%%%%%%%%%%%%%%%%%%%%%%%%%%%%%%%%%%%%%%%%%%%%%%%%%%%%%%%%%%%%%%%%%%%%%%%%%%%%%
%%%%%%%%%%%%%%%%%%%%%%%%%%%%%%%%%%%%%%%%%%%%%%%%%%%%%%%%%%%%%%%%%%%%%%%%%%%%%%%
% APPENDIX
%%%%%%%%%%%%%%%%%%%%%%%%%%%%%%%%%%%%%%%%%%%%%%%%%%%%%%%%%%%%%%%%%%%%%%%%%%%%%%%
%%%%%%%%%%%%%%%%%%%%%%%%%%%%%%%%%%%%%%%%%%%%%%%%%%%%%%%%%%%%%%%%%%%%%%%%%%%%%%%
\newpage
\appendix
\onecolumn
\section{Conditional Generation Figures}
\begin{figure*}[h]
\begin{center}
\subfigure[$Real$]{
\begin{minipage}[t]{0.33\linewidth}
\includegraphics[width=0.9\linewidth]{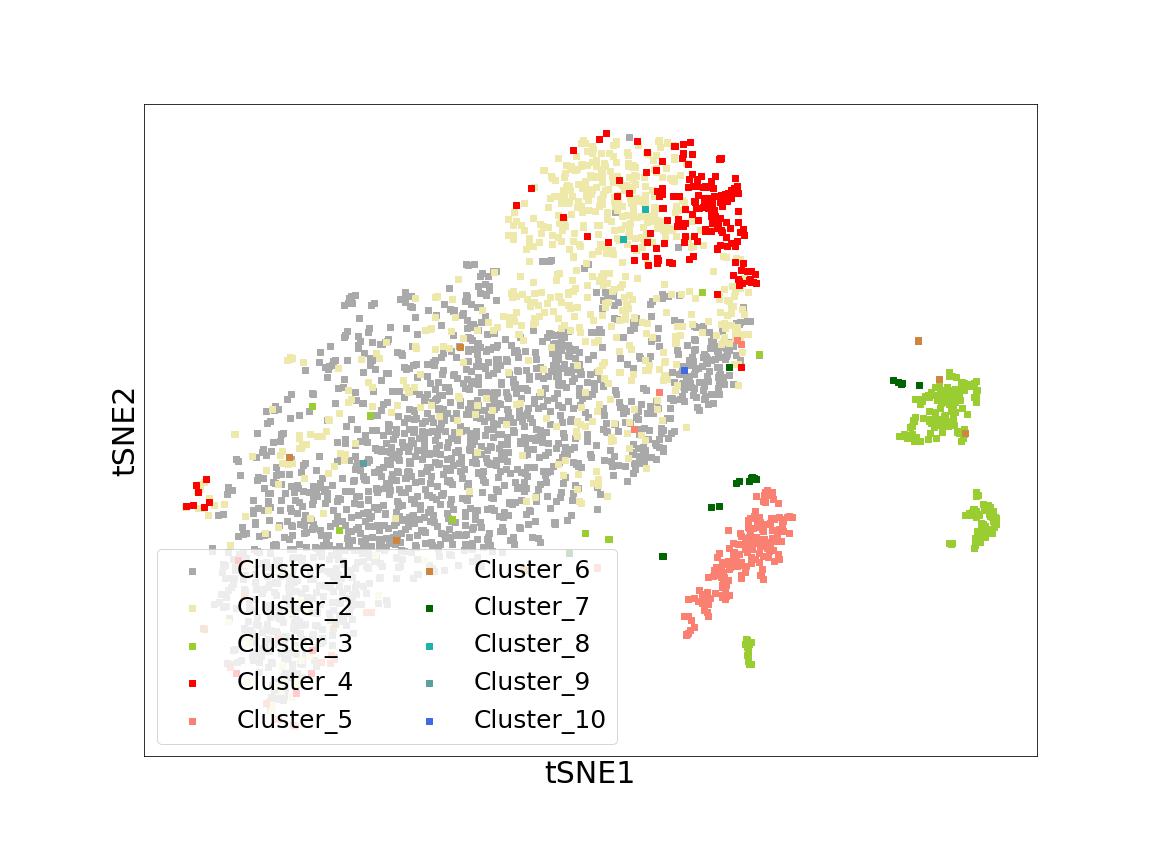}
%\caption{fig1}
\centering
\end{minipage}%
}%
\subfigure[$\omega:-1, c:4,samples:3k$]{
\begin{minipage}[t]{0.33\linewidth}
\includegraphics[width=0.9\linewidth]{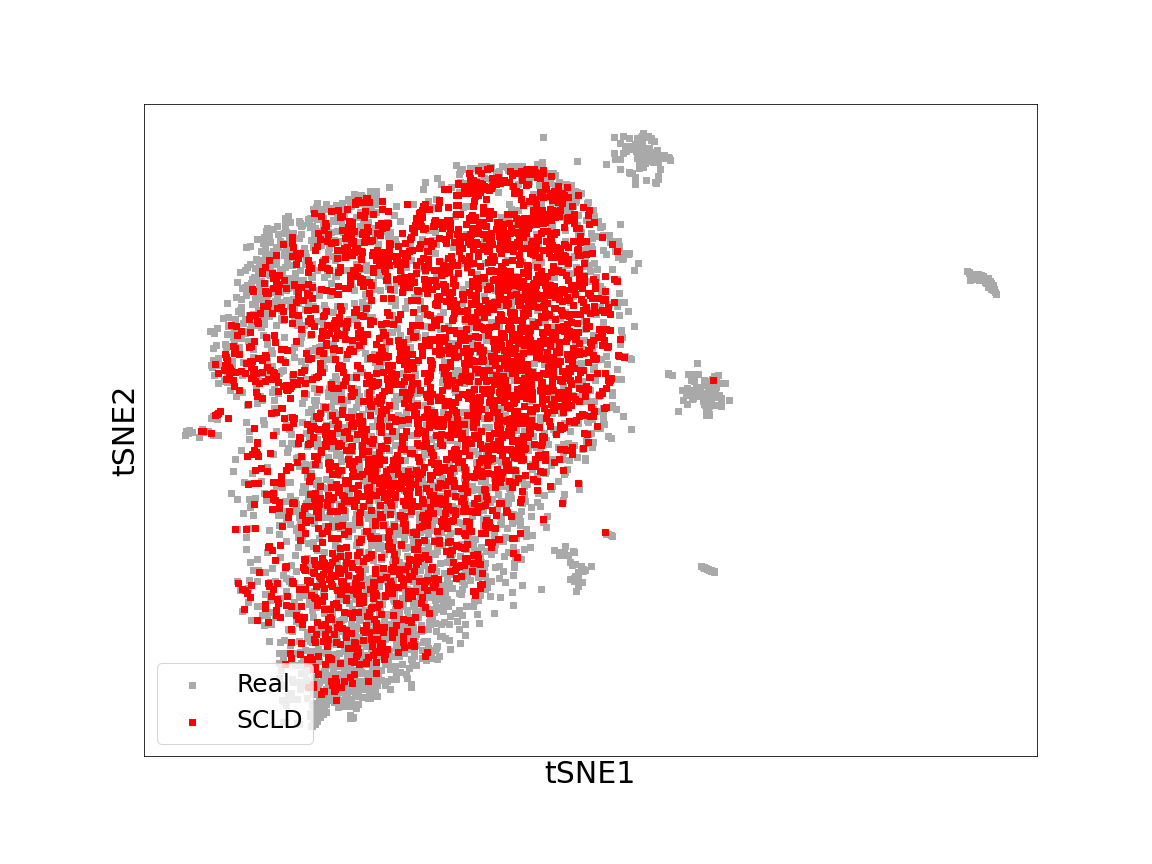}
%\caption{fig1}
\centering
\end{minipage}%
}%
\subfigure[$\omega:-0.5, c:4,samples:2.5k$]{
\begin{minipage}[t]{0.33\linewidth}
\includegraphics[width=0.9\linewidth]{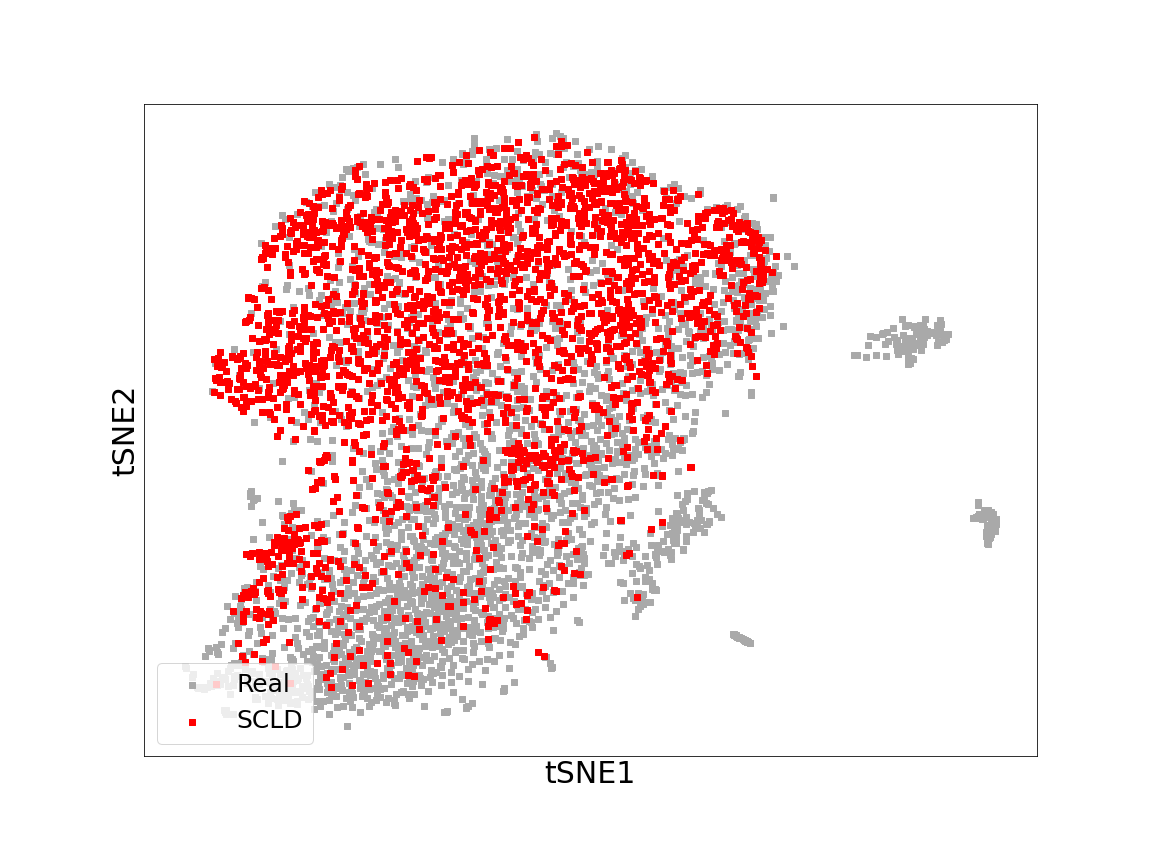}
%\caption{fig1}
\centering
\end{minipage}%
}%
\vspace{-1.2em}
\subfigure[$\omega:0, c:4,samples:2k$]{
\begin{minipage}[t]{0.33\linewidth}
\centering
\includegraphics[width=0.9\linewidth]{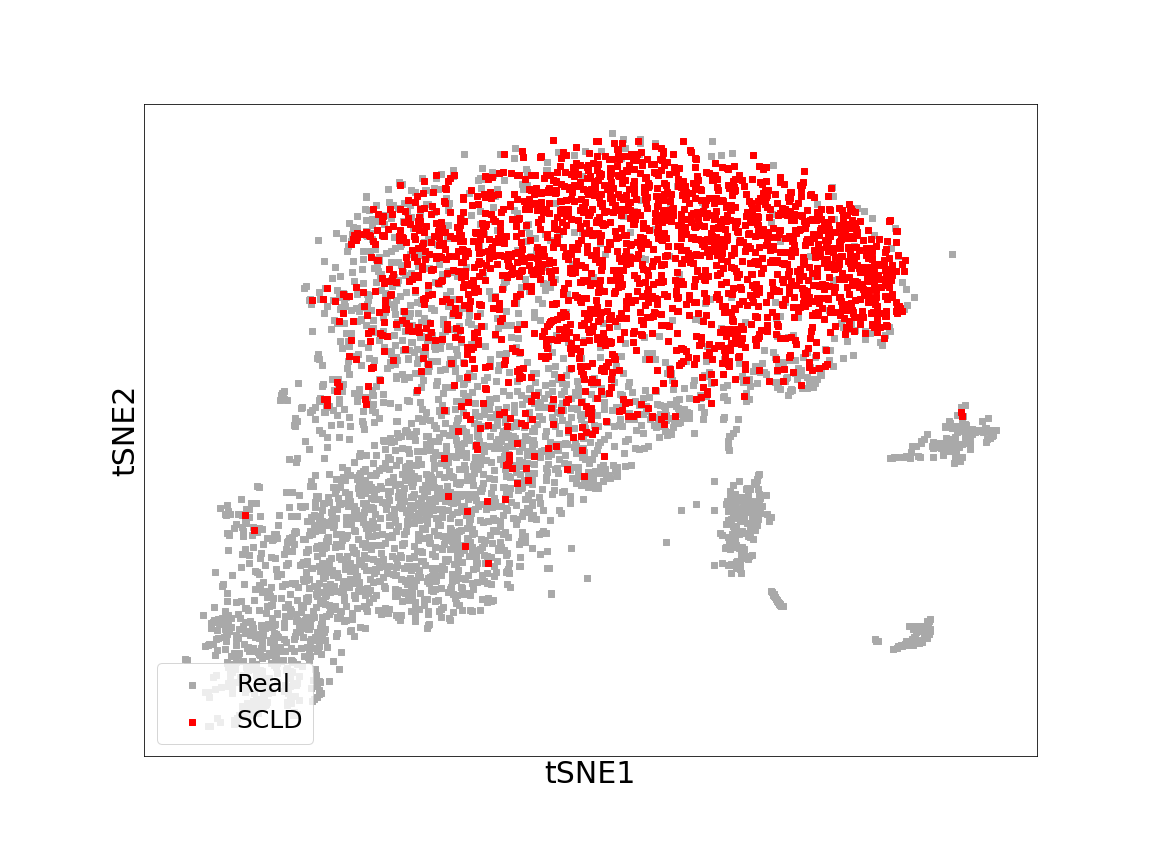}
%\caption{fig2}
\end{minipage}%
}%
\subfigure[$\omega:0.5, c:4,samples:1.5k$]{
\begin{minipage}[t]{0.33\linewidth}
\centering
\includegraphics[width=0.9\linewidth]{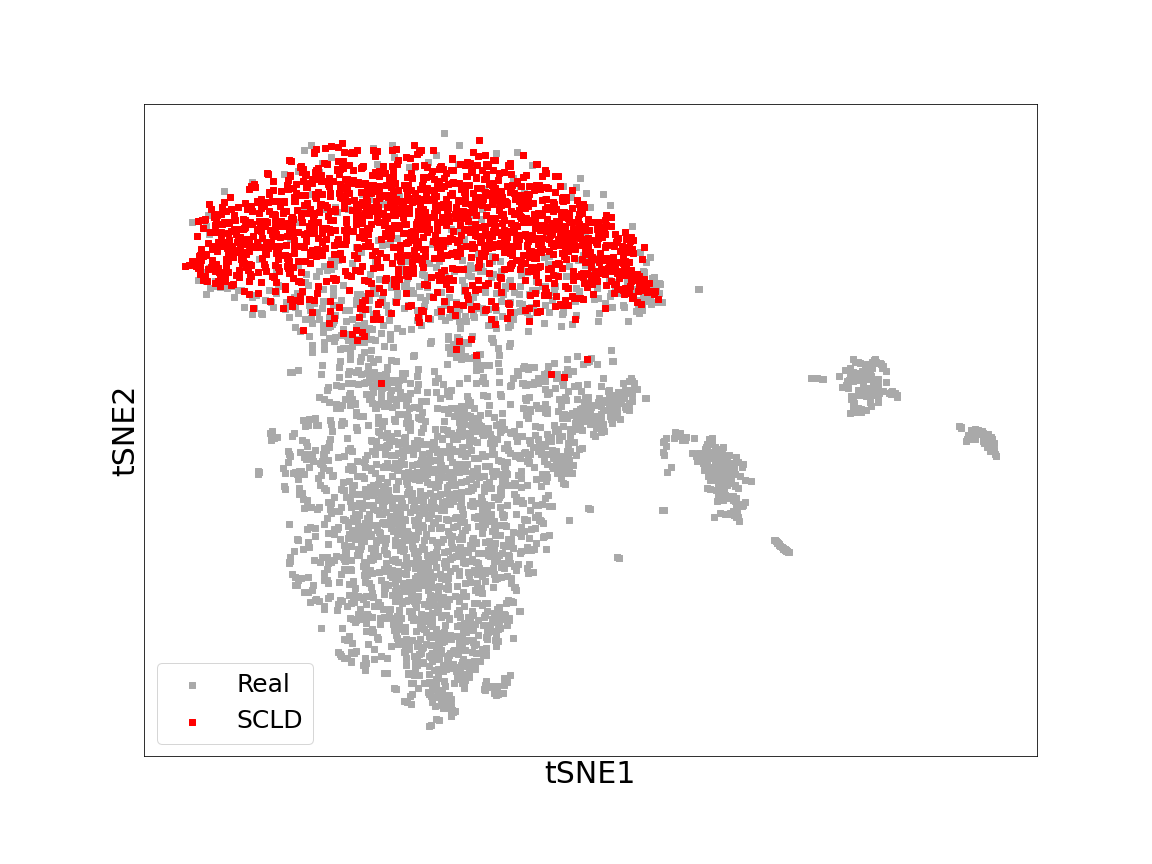}
%\caption{fig1}
\end{minipage}%
}%
\subfigure[$\omega:1, c:4, samples:1k$]{
\begin{minipage}[t]{0.33\linewidth}
\centering
\includegraphics[width=0.9\linewidth]{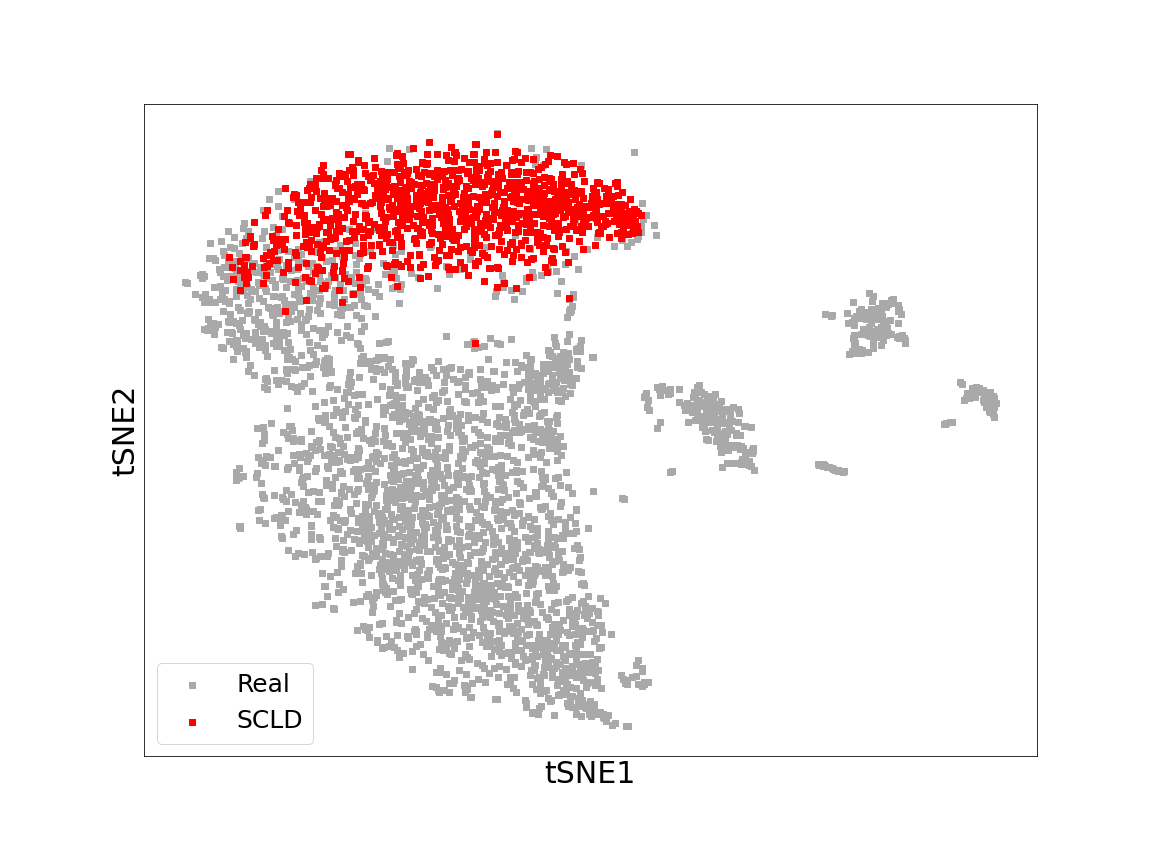}
%\caption{fig2}
\end{minipage}%
}%

\subfigure[$Real$]{
\begin{minipage}[t]{0.33\linewidth}
\centering
\includegraphics[width=0.9\linewidth]{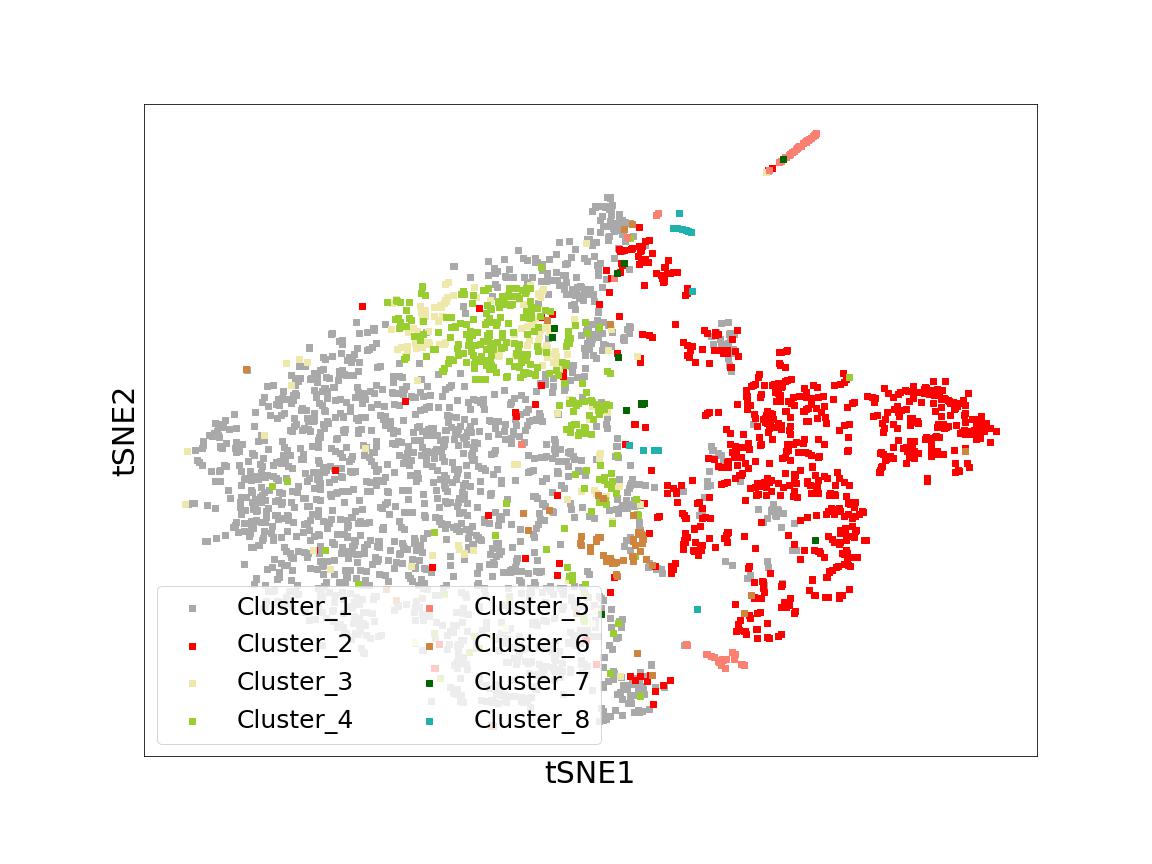}
%\caption{fig1}
\end{minipage}%
}%
\subfigure[$\omega:-1,c:2, samples:3k$]{
\begin{minipage}[t]{0.33\linewidth}
\centering
\includegraphics[width=0.9\linewidth]{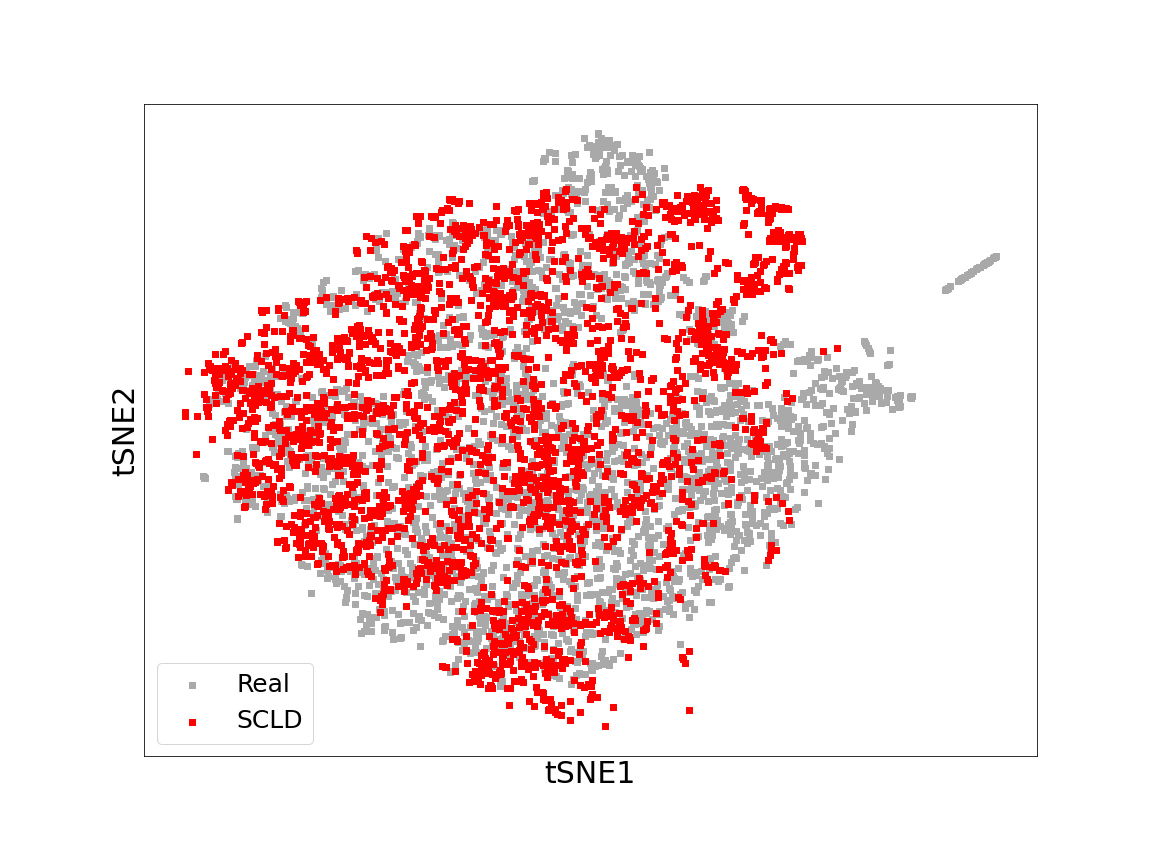}
%\caption{fig1}
\end{minipage}%
}%
\subfigure[$\omega:-0.5,c:2, samples:2.5k$]{
\begin{minipage}[t]{0.33\linewidth}
\centering
\includegraphics[width=0.9\linewidth]{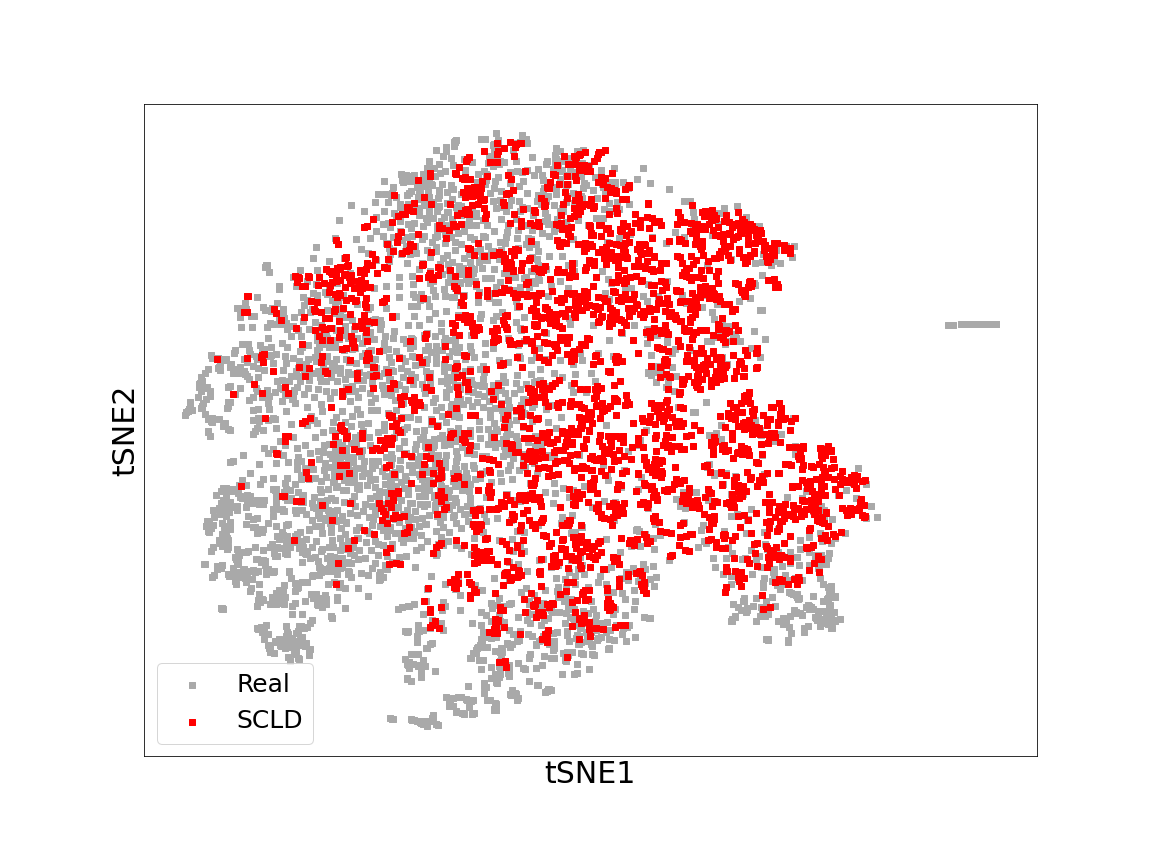}
%\caption{fig1}
\end{minipage}%
}%
\vspace{-1.2em}
\subfigure[$\omega:0,c:2, samples:2k$]{
\begin{minipage}[t]{0.33\linewidth}
\centering
\includegraphics[width=0.9\linewidth]{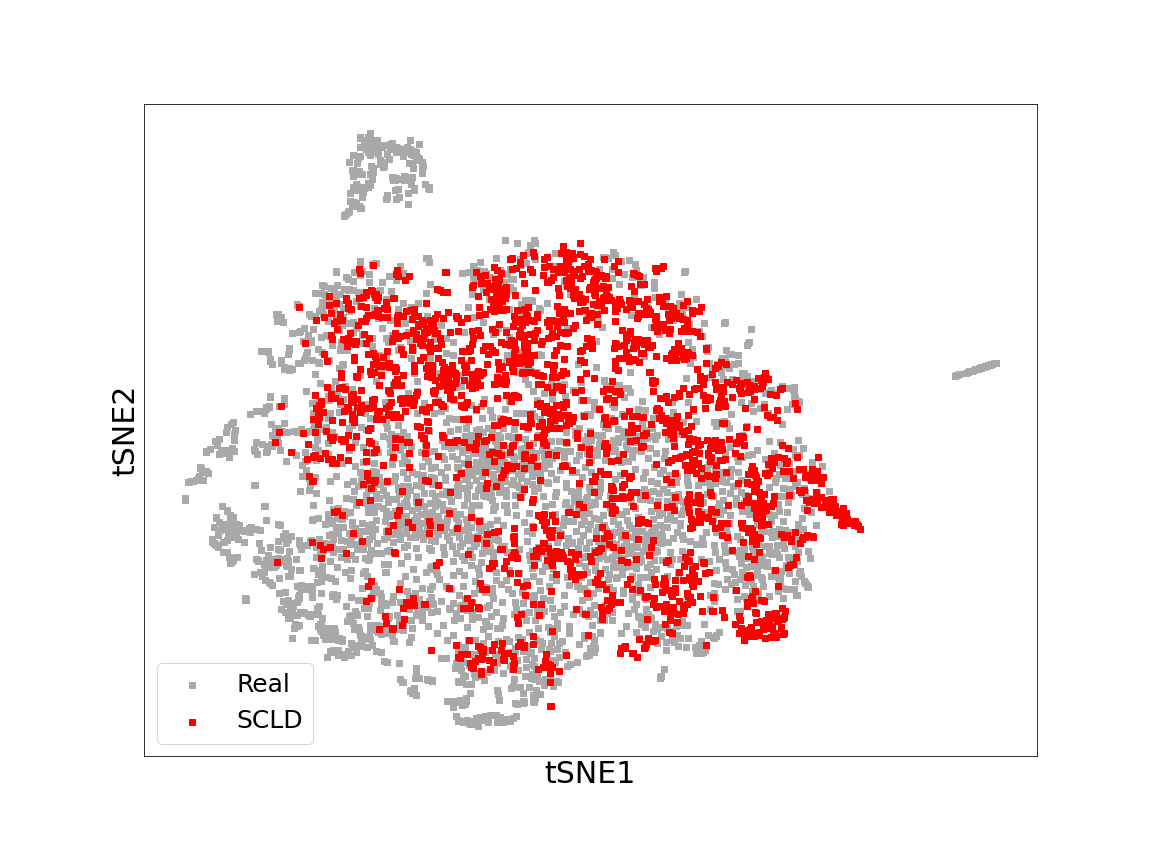}
%\caption{fig2}
\end{minipage}%
}%
\subfigure[$\omega:0.5, c:2, samples:1.5k$]{
\begin{minipage}[t]{0.33\linewidth}
\centering
\includegraphics[width=0.9\linewidth]{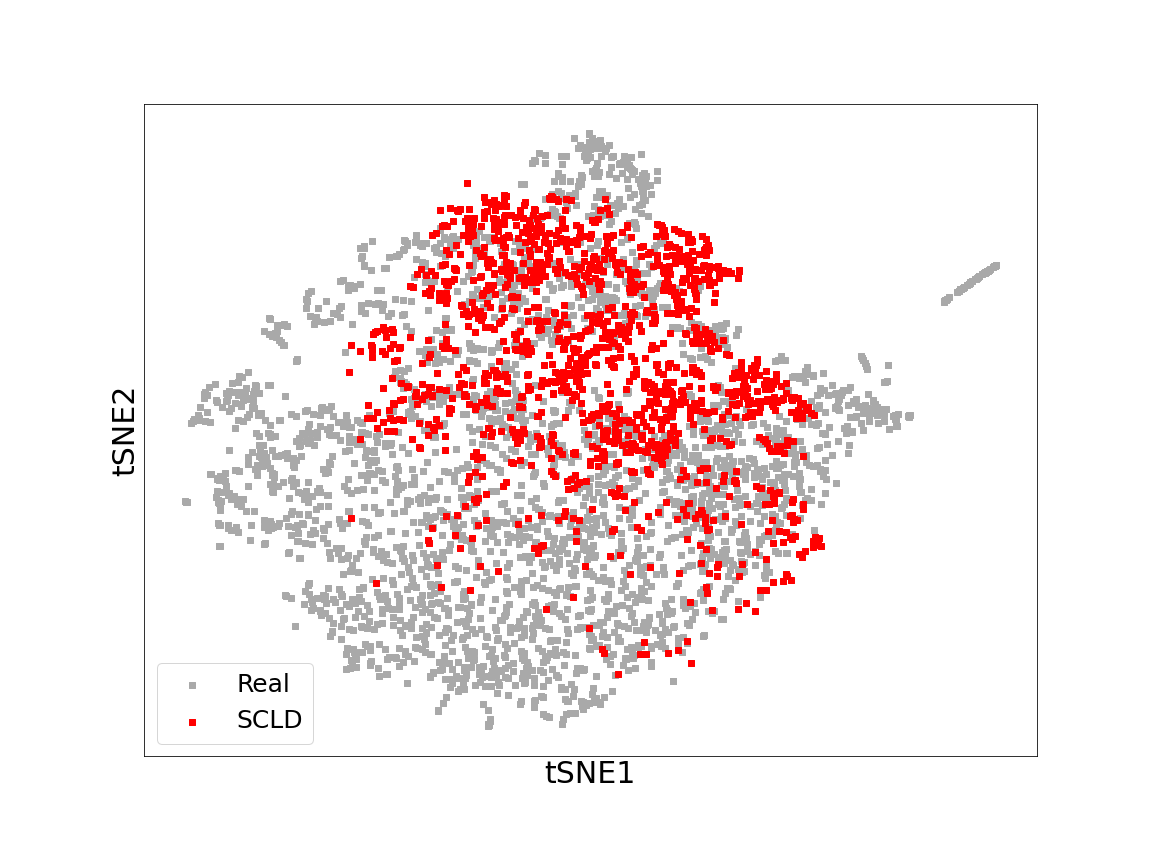}
%\caption{fig1}
\end{minipage}%
}%
\subfigure[$\omega:1, c:2, samples:1k$]{
\begin{minipage}[t]{0.33\linewidth}
\centering
\includegraphics[width=0.9\linewidth]{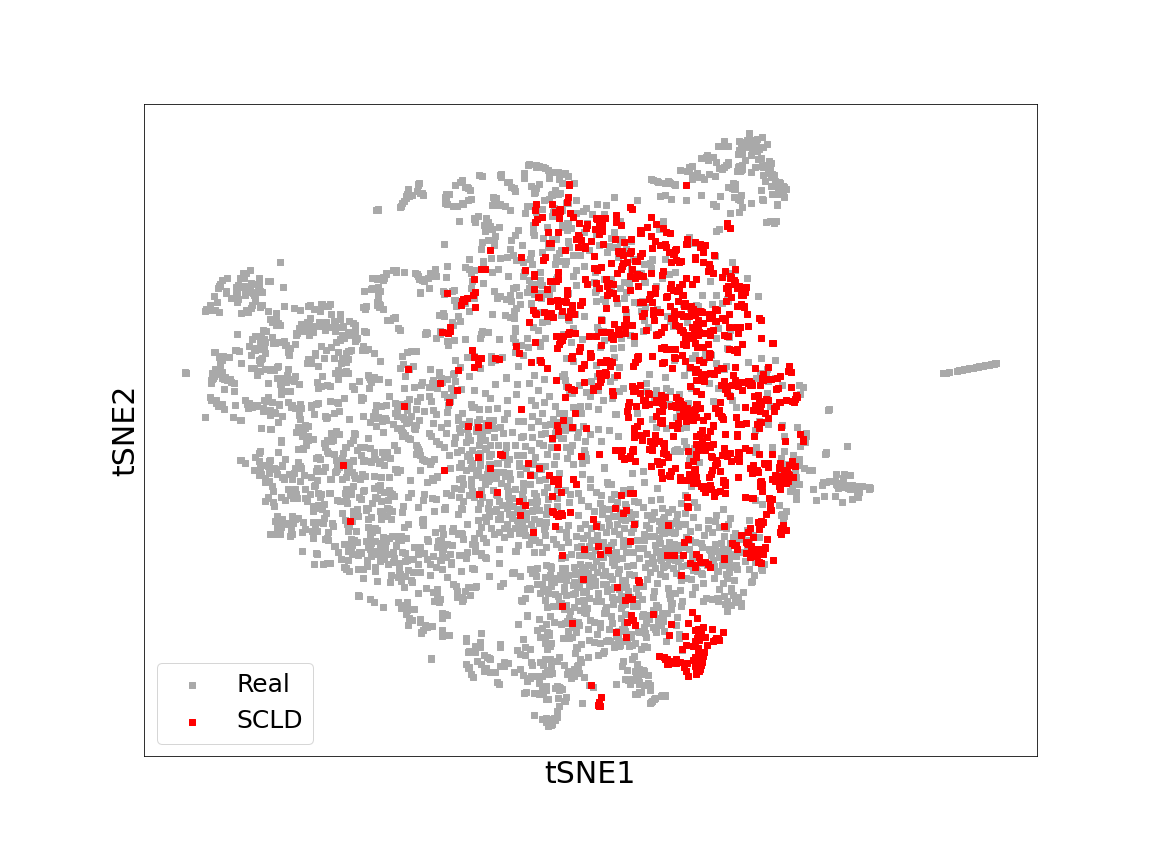}
%\caption{fig2}
\end{minipage}%
}%
\caption{ Evaluation of the conditional generation of PBMC(a-f) and Brain Small(g-l) cells. Setting $\omega$ as -1, 0, and 1 demonstrated the classifier-guidance capacity of the SCLD. Using real cells as a reference, it is easier to figure out how the SCLD changes from "holistic" to specific cellular subpopulations.} 

\label{fig:test}

\end{center}
\end{figure*}

A visual picture of conditional control is shown here (Figure \ref{fig:test}), and related analysis can be viewed in the text.

 \section{Learning \& Training}
In this work, we propose a pre-guidance mechanism for generating synthetic single-cell sequencing data using a diffusion model guided by category labels and global information, as well as a post-guidance mechanism that employs a $Scorer$ to improve the quality of generated cell subpopulations. Latent Space Guided Diffusion can train the SCLD more effectively by mapping single-cell scRNA-seq samples into a continuous latent space. The complete process of the SCLD consists of a training algorithm and a sampling algorithm, which are outlined in Algorithm \ref{alg:algorithm1} and Algorithm \ref{alg:algorithm2}. The training algorithm comprises category information, sampling of Gaussian noise, and computation of the loss function. The sampling algorithm involves class information parameters, controllable generation of scRNA-seq samples, and mapping of the scRNA-seq samples from the latent space back to the high-dimensional space. It is worth mentioning that a storage queue stores the continuously generated scRNA-seq samples produced by the diffusion model to avoid the computational overhead of separating the diffusion model and $Scorer$. By taking advantage of the queue's characteristics, the diffusion model and the $Scorer$ can run synchronously rather than waiting for the cells to be entirely generated by the diffusion model before scoring.

The method for generating synthetic single-cell sequencing data involves a diffusion model guided by category labels and global information. It includes a pre-guidance mechanism to incorporate category information and a post-guidance mechanism to improve the quality of generated cells. The process also entails mapping single-cell data into a continuous latent space and training the diffusion model using a training algorithm and a sampling algorithm. These steps are outlined in Algorithm \ref{alg:algorithm1} and Algorithm \ref{alg:algorithm2}. The SCLD offers a solution for generating synthetic single-cell sequencing data that accurately reflects real-world data.

\begin{minipage}{.5\linewidth}
\begin{algorithm}[H]
    \caption{Training conditional diffusion model}
    \label{alg:algorithm1}
    \textbf{Input}: $X$: scRNA-seq samples data, \\
    $Labels$: Category embedding\\
    \textbf{Output}:$loss$:loss function\\
    \textbf{Require}: $k$: Size of global training
    \begin{algorithmic}[1] %[1] enables line numbers
        \REPEAT
        \STATE $x_0, l = sample(X, Labels);$ 
        \STATE $z_0 = Round(x_0);$ 
        \STATE $l[:k] = global;$ 
        \STATE $t \sim Uniform(\{1,...,T\});$
        \STATE $\epsilon \sim \mathcal{N}(0, I);$ 
        \STATE $z_t = \sqrt{\Bar{\alpha}_t}z_0 + \sqrt{1-\Bar{\alpha}_t}\epsilon;$
        \STATE $loss=||\epsilon_\theta(z_t, l, t)-\epsilon||^2;$
        \UNTIL{converged}
    \end{algorithmic}
\end{algorithm}
\end{minipage}%
\begin{minipage}{.5\linewidth}
\begin{algorithm}[H]
    \caption{Sample with the SCLD}
    \label{alg:algorithm2}
    \textbf{Input}: $x_T$: standard normal distribution,\\
    \textbf{Output}: $x_0$: scRNA-seq Sample generated\\
    \textbf{Require}: $c$: Conditional information for sampling\\
    $\omega$: Conditionally controllable factor
    \begin{algorithmic}[1] %[1] enables line numbers
        \STATE $z_T = Round(x_T)$
        \FOR{$t=T,...,1$}
        \STATE $n \sim \mathcal{N}(0, I)$ if t $>$ 1, else n=0;
        \STATE $\Tilde{\epsilon}_t = (1+\omega)\epsilon_\theta(z_t, c)-\omega\epsilon_\theta(z_t);$
        \STATE $z_{t-1} = \frac{1}{\sqrt{\alpha_t}}(z_t - \frac{1-\alpha_t}{\sqrt{1-\alpha_t}}{\Tilde{\epsilon}}_t(z_t, t)) + \sigma_tn;$
        \ENDFOR
        \STATE $x_0 = Map(z_0);$
        \STATE \textbf{return} $x_0;$
    \end{algorithmic}
\end{algorithm}
\end{minipage}

\section{Experiments Details and Hyperparameters}
% The SCLD follows the backbone of DDPM, which is a U-Net based on a Wide ResNet \cite{resnet}. In order to reduce the training costs as much as possible and ensure that the performance does not drop significantly, the SCLD halves all the parameters of the original model. In follow-up experiments, the performance of the SCLD is not significantly affected.

% Firstly we set $T$ = 1000 without a sweep, and chose a linear schedule from $\beta_1$ = $10^{-4}$ to $\beta_2$ = $0.02$. Then we set the dropout rate to 0.1 by sweeping over the values $\{0.1, 0.2, 0.3, 0.4\}$. What's more, the batchsize we set is 128 and the optimizer we use is AdamW \cite{adamW} with default parameters. The experiments are run in the Linux environment by PyTorch \cite{pytorch}. The graphics card platform is a single Tesla V100 (similar to prior work).

In our experiments, the Single-Cell Latent Diffusion (SCLD) model is built upon the backbone of the Denoising Diffusion Probabilistic Model (DDPM), which utilizes a U-Net architecture based on a Wide Residual Network (Wide ResNet) \cite{resnet}. To reduce the training costs while maintaining the performance of the model, we make modifications to the original DDPM architecture by halving all the parameters. In subsequent experiments, we observe that the performance of the SCLD model remains largely unaffected by these changes.

For the experimental setup, we initially set the total number of time steps, $T$, to 1000 without performing a parameter sweep. We adopt a linear schedule for the noise schedule parameter, $\beta$, with values ranging from $\beta_1 = 10^{-4}$ to $\beta_2 = 0.02$. To determine the optimal dropout rate, we perform a sweep over the values ${0.1, 0.2, 0.3, 0.4}$ and eventually select a dropout rate of 0.1, which provides the best performance.
Furthermore, we use a batch size of 128 and total step of 20000 during training, and employ the AdamW optimizer \cite{adamW} with its default parameters to update the model weights. Our experiments are conducted in a Linux environment using the PyTorch deep learning framework \cite{pytorch}. For hardware, we rely on a single Tesla V100 graphics processing unit (GPU), which is consistent with the hardware used in prior related works.

This experimental setup, including the architectural modifications and parameter selections, allows us to effectively evaluate the performance of the SCLD model in analyzing single-cell sequencing data while minimizing training costs. Our results demonstrate the potential of the SCLD model as an efficient and effective solution for single-cell sequencing data analysis, with the ability to maintain its performance even when parameters are reduced.

%%%%%%%%%%%%%%%%%%%%%%%%%%%%%%%%%%%%%%%%%%%%%%%%%%%%%%%%%%%%%%%%%%%%%%%%%%%%%%%
%%%%%%%%%%%%%%%%%%%%%%%%%%%%%%%%%%%%%%%%%%%%%%%%%%%%%%%%%%%%%%%%%%%%%%%%%%%%%%%

\end{document}